\documentclass[twocolumn]{revtex4}

\begin{document}

\title{Relations between cloning and the universal NOT 
from conservation laws}
\author{S. J. van Enk}
\affiliation{Bell Labs, Lucent Technologies\\
600-700 Mountain Ave, Murray Hill, NJ 07974}
\date{\today}

\begin{abstract}
We discuss certain relations between cloning and the NOT operation
that can be derived from conservation laws alone.
Those relations link the limitations on 
cloning and the NOT operation possibly
imposed by {\em other} laws of Nature.
Our result is quite general and holds both in classical and quantum-mechanical
worlds, for both optimal and suboptimal
operations, and for bosons
as well as fermions.
\end{abstract}

\maketitle
As is well known,
there are fundamental limitations on the accuracy of
certain quantum operations,
with cloning and the 
NOT operation applied to quantum bits being two prime 
examples \cite{nocloning}--\cite{martini}.
Those
limitations
are independent of physical implementation. 
For instance, it is irrelevant whether the qubits are implemented 
using Josephson junctions, ions, or photons. Similarly, 
it is irrelevant whether the two basis states $|0\rangle$ and $|1\rangle$
of the qubit correspond to 
eigenstates of different charge, of different energy, or of
different angular momentum.
Once one has chosen a particular implementation, however,
there are often, if not always, conservation laws 
that must be obeyed.
Indeed, there must be at least one physical 
quantity that takes different values in the states
$|0\rangle$ and $|1\rangle$, otherwise the two states would not be
orthogonal and distinguishable. Depending on the situation there will be
a conservation law for that quantity, or for the
complementary variable, or for both. For example, although
there are no conservation laws for position or time, there are for
linear momentum and energy.

In the following we consider relations between limitations on 
cloning and the NOT operation 
that arise from two simple assumptions:
\begin{enumerate}
\item The states $|0\rangle$ and $|1\rangle$ correspond to
eigenstates of some conserved quantity with eigenvalues $-1$ and $+1$
in some appropriate unit. For concreteness
we will say that the states are eigenstates of 
``angular momentum''.
\item Each qubit is a ``particle'' 
and a conservation law
holds for the number of particles. 
\end{enumerate}
In relation to assumption 2, note that in certain contexts
(when considering atoms, ions, quantum dots, or any other material entities
as qubits)
conservation of particle number is appropriate, whereas
in other contexts (for example, when one considers photons as qubits 
\cite{simon,bouwmeester, martini}) conservation
of excitation number is more appropriate. 
In the following considerations these two cases 
are mathematically
equivalent,
and for ease of notation we will
henceforth refer to particles and use the particle conservation law.

It is important to note that
in spite of the quantum-mechanical notation and terminology
used here,
the assumptions just mentioned by themselves make no use of quantum mechanics.
In particular, here and in the following we will only need to discuss
particles in states $|0\rangle$ and $|1\rangle$, but not in superpositions
of $|0\rangle$ and $|1\rangle$.
The relations we will find between cloning and NOT operations
hold, therefore, just as well
for classical cloning procedures and NOT operations. 
But since in the end we are mostly interested in understanding the
quantum-mechanical results, we adopt quantum-mechanical notation.

Suppose we start out with $N$ particles in the state $|0\rangle$
and attempt to generate $M>N$ clones in the same state.
In general, 
we will end up not only with $M$ clones 
in states $|0\rangle$ and possibly in state $|1\rangle$, but, by assumption 1,
with some nonzero number $K$ (to be determined later) of ancilla particles 
that must be present to compensate for the
amount of angular momentum produced or destroyed in the cloning process.
By assumption 2 then, we must have borrowed $M+K-N$ particles 
from elsewhere, 
a ``reservoir'' of particles. We assume the reservoir starts out
in a state with equal numbers of particles, say $L$,
in states $|0\rangle$ and $|1\rangle$. 
Thus, we denote the initial state by
\begin{equation}
|N,0\rangle\otimes |L,L\rangle
\end{equation}
where $|n,m\rangle$ denotes a state with $n$ particles in state $|0\rangle$
and $m$ particles in state $|1\rangle$.
The attempted cloning operation may then be described by 
the transformation
\begin{equation}\label{clone}
|N,0\rangle\otimes|L,L\rangle\mapsto
\sum_{a,b}A_{a,b}|a,M-a\rangle\otimes|b,K-b\rangle\otimes|L',L'\rangle.
\end{equation}
The first ket on the right-hand side refers to clones, the second ket
to ancillas, and the third ket to the reservoir.
The coefficients
$A_{a,b}$ determine
in some unspecified way the probabilities $p_{a,b}$ to 
find $a$ clones in state $|0\rangle$ and $b$ ancillas in state $|0\rangle$.
In quantum mechanics we would have
$p_{a,b}=|A_{a,b}|^2$. Although we wrote down a quantum-mechanical 
superposition we may just as well regard  
the superposition as a classical probability distribution over the
various possible outcomes of the cloning operation.
The number $L'=L+(N-M-K)/2$ is fixed by particle number conservation,
and $N-M-K$ must be an even number.

Assumption 1 puts a constraint
on the numbers $a,b$.
Angular momentum conservation gives
\begin{equation}\label{c1}
2(a+b)=N+K+M.
\end{equation} 
At this point $K$, the number of ancillas, is still somewhat arbitrary.
Namely, after having fixed the cloning operation we can 
always {\em increase} the number of ancillas by taking 
some (even) number of particles from the reservoir
and promote them 
to ancillas. This does not affect the cloning operation
and, provided
we transfer equal numbers of particles in
states $|0\rangle$ and $|1\rangle$ from the reservoir,
does not affect angular momentum conservation of clones and ancillas either.
It makes sense then to take the {\em smallest} possible number $K$
consistent with all conservation laws as the canonical
number of ancillas.
That minimum number is easy to determine from constraint (\ref{c1}):
suppose Nature allows us at least sometimes to
achieve perfect cloning of the state $|0\rangle$ with some nonzero 
probability. In that case we have a term in the superposition with 
$a=M$. The minimum $b$ allowed is, of course, $b=0$,
so that the minimum $K$ consistent with (\ref{c1}) is
\begin{equation}
K=M-N.
\end{equation}
Adopting this value for $K$ then fixes $b$ to be
\begin{equation}
b=M-a.
\end{equation}
With these ingredients, the cloning operation
(\ref{clone}) can be rewritten as
\begin{eqnarray}\label{clone2}
&&|N,0\rangle\otimes|L,L\rangle\mapsto\nonumber\\
&&\sum_{a}A_{a,M-a}|a,M-a\rangle\otimes|M-a,a-N\rangle\otimes|L',L'\rangle.
\end{eqnarray}
The average cloning fidelity $F_{{\rm clone}}$ may be defined 
as
\begin{equation}
F_{{\rm clone}}=\sum_a p_a\frac{a}{M},
\end{equation} 
where $p_a$ (determined by $A_{a,M-a}$)
gives the probability to find $a$ clones in the (correct)
state $|0\rangle$, and where the cloning fidelity for a state with $a$
clones out of $M$ in the correct state is defined to be the ratio $a/M$.
This in fact corresponds to the standard definition
of cloning fidelity, see Ref.~\cite{howell} and the discussion
below. 

The ancillas compensate for the angular momentum
produced in the cloning procedure and thus, roughly speaking, they will
end up in a state with
angular momentum opposite to that of the clones. 
But ``flipping the angular momentum''
of a qubit state is the same as applying the NOT operation \cite{gisinp}.
Thus, the better
the cloning procedure works, the better
the NOT operation will be implemented on the ancillas. 
This is why there is a strong connection between the 
fidelity $F_{{\rm clone}}$ of the cloning operation
and a similar fidelity $F_{{\rm NOT}}$ one can define 
for the NOT operation. 
Namely, $F_{{\rm NOT}}$ is analogously defined as the average
of the ratios of the number
of ancillas in state $|1\rangle$, $a-N$, 
and the total number of ancillas, $M-N$. Thus,
\begin{equation}
F_{{\rm NOT}}=\sum_a p_a\frac{a-N}{M-N}.
\end{equation} 
But this immediately gives us a relation
between $F_{{\rm clone}}$ and $F_{{\rm NOT}}$ that is {\em 
independent} of 
the values of $p_a$ \footnote{Strictly speaking,
in the derivation of the result we assumed
the value of $p_a$ to be nonzero for $a=M$. However, even if
$p_a=0$ the result (\ref{clonot}) still holds provided we keep using
$M-N$ as the value for $K$.}
\begin{equation}\label{clonot}
(M-N)F_{{\rm NOT}}=MF_{{\rm clone}}-N.
\end{equation}
So far, we only considered cloning and the NOT as applied to
$|0\rangle$. But for universal cloners and the universal-NOT operation
the fidelities are, by definition, independent
of the input state. Thus, the relation (\ref{clonot})
holds for any (optimal or suboptimal)
universal cloner and universal-NOT operation.
The independence of that relation on 
the details of the transformation (\ref{clone2}) demonstrates
the generality of the result.

Instead of starting out, 
as we did here,
with an operation that is supposed to clone the state $|0\rangle$,
we might as well have begun with describing an operation that is
supposed to apply a NOT operation. 
The constraints
we would find then are exactly the same as we found before.
And so we would find the same relation (\ref{clonot}) again,
even if we had found different coefficients
$B_{a,b}$ instead of $A_{a,b}$.
This, combined with the simple linear relationship between
$F_{{\rm clone}}$ and $F_{{\rm NOT}}$,
implies in particular
that optimizing the NOT operation would automatically
optimize
the cloning operation, and {\em vice versa}. 

The relation (\ref{clonot}) quantifies to what extent
the NOT operation can be performed given how well cloning can be performed,
and {\em vice versa}. For example, if one can perform one perfectly, the other
procedure can be performed perfectly as well. 
That, of course, reflects what is possible in a classical world,
but also what is possible quantum-mechanically
when one knows the input state.
From (\ref{clonot}) we see that in general $F_{{\rm NOT}}$ is never larger
than $F_{{\rm clone}}$, but in the limit of 
$M\rightarrow\infty$ with $N$ finite one gets
$F_{{\rm NOT}}=F_{{\rm clone}}$.
The optimum quantum cloning fidelity for universal
cloning of arbitrary unknown input states
is well-known to be \cite{gisinm} 
\begin{equation}
F^{{\rm opt}}_{{\rm clone}}=\frac{M(N+1)+N}{M(N+2)}.
\end{equation}
This combined with Eq.~(\ref{clonot}) immediately
yields the optimum universal-NOT fidelity:
\begin{equation}
F^{{\rm opt}}_{{\rm NOT}}=\frac{N+1}{N+2},
\end{equation}
which turns out to be independent of $M$.
And indeed, this is identical to the result
obtained in \cite{not,gisinp} by other means.

The above-used notation is appropriate for bosons, with $a,b$
being occupation numbers of certain ``modes''. Nevertheless,
the results are equally valid for fermions.
Indeed, there are in fact {\em two} different interpretations
of ``cloning'' when it is applied to bosons, and one of those interpretations
applies to fermions as well.
For simplicity, first consider $1\rightarrow2$ cloning.
The (arbitrary, unknown) state to be cloned can be written as
$A|0\rangle+B|1\rangle$,
and cloning, in the standard terminology, 
would correspond to the transformation
\begin{equation}\label{T1}
A|0\rangle+B|1\rangle\mapsto
[A|0\rangle+B|1\rangle]^{\otimes 2},
\end{equation}
which adds a second particle and a second system.
This formulation works for fermions as well as for bosons.
Alternatively, we may write the same state in terms of creation
operators $C^{\dagger}_{0,1}$
for excitations in the different modes 
$0$
and $1$. For bosons, but not for fermions, it is possible
to create more than one excitation in a single mode.
Creating two bosons in the same mode 
starting from a 
single boson may also be considered a form of cloning. This
would correspond to the transformation
\begin{equation}\label{T2}
[AC^{\dagger}_0+BC^{\dagger}_1]|{\rm vacuum}\rangle
\mapsto
\frac{[AC^{\dagger}_0+BC^{\dagger}_1]^2}{\sqrt{2}}|{\rm vacuum}\rangle,
\end{equation}
which adds a second particle (excitation) 
but does not add a second system (mode).
Thus, although the initial states in the transformations (\ref{T1})
and (\ref{T2}) are the same, the final states are different.
However, it is easy to verify there is a unitary operation
taking one final state
to the other. In other words, the two descriptions are unitarily
equivalent. It is also easy to check that the fidelity in terms of 
occupation numbers of modes used in the present paper 
(corresponding to cloning in the sense of (\ref{T2})) is equivalent
to the usual definition of fidelity provided
the cloning operation is symmetric (i.e., all clones in Eq.~(\ref{T1})
end up in the same state). This very same point was made in
Ref.~\cite{howell}, where details on the equivalence of
the two definitions of fidelity can be found.
Hence, the limits on
$1\rightarrow2$ cloning are in fact exactly the same 
irrespective of which definition of cloning one prefers.

More generally, one can show 
the two different final states that appear
in general $N\rightarrow M$ cloning procedures are unitarily equivalent,
and that the two corresponding definitions of fidelity are identical
for symmetric cloners.
In addition, in a similar manner
one may
use two different definitions and descriptions for the NOT operation
acting on bosons. 
In the end those two formulations, too,
are equivalent, with one of them applicable to fermions.

The {\em optimal} universal cloner applied to the polarization degree
of freedom of photons can be and has been implemented 
using stimulated emission 
\cite{simon,bouwmeester,martini}.
In that context, the operation (\ref{clone2}) can be understood as follows:
the initial state consists of $N$ photons in a particular
spatial mode (denoted by '1'), all $\sigma^-$ polarized.
The ``reservoir'' consists of $2L$ excited atoms: for instance, if 
one uses a $J=1/2\rightarrow J'=1/2$ transition, then
an atom in the excited state $|J'_z=\pm 1/2\rangle$ ``stores'' 
a $\sigma^{\pm}$-polarized photon.
By stimulated emission
one produces with some probability  
$M$ photons in the same spatial mode '1' (the clones),
and $M-N$ photons
in a different spatial mode '2' (the ancillas). The total number of 
atomic and photonic excitations
is conserved, i.e., $2L-2L'=2M-2N$ atoms 
have decayed to the appropriate ground states. Obviously, 
angular momentum is conserved too in this case, 
as expressed by selection rules. The unitary operation
implementing the optimal cloner and the optimal NOT
corresponding to this particular physical implementation is given 
in Refs.~\cite{simon,bouwmeester,martini}.

In conclusion then,
we showed a strong relation exists
between (universal) cloning and the (universal) NOT operation. 
That a relation exists between the {\em optimum} fidelities
for quantum cloning and the quantum universal NOT operations
had been noticed before in the context of photons, as we just mentioned, 
in Refs.~\cite{simon,bouwmeester,martini}.
But here
we demonstrated that the relation (\ref{clonot}) holds more generally:
it holds for suboptimal procedures, it holds for fermions as well as bosons,
and it holds in the classical world. Moreover, 
we explained {\em why}
an optimum cloner also implements the optimal NOT 
operation.
The only assumptions needed to derive this were 
simple conservation laws.
Conversely, it had been noted before that for the optimal cloner
a conservation law holds: a particularly nice form
of such a conservation law can be found in Ref.~\cite{calsa}.

Finally, we note that the impossibility of the {\em perfect}
NOT operation
arises from it not being a completely positive map \cite{not},
whereas no-cloning arises from the linearity of quantum mechanics 
\cite{nocloning}. On the other hand,
the optimum fidelities of the corresponding {\em imperfect} quantum
operations
are determined by the unitarity of quantum mechanics.
Eq.~(\ref{clonot}), however, uses none of those properties: 
Unitarity or linearity or complete
positivity put restrictions on the values
of the coefficients $A_{a,b}$ in Eq.~(\ref{clone}), 
but relation (\ref{clonot})
holds independent of the precise values of $A_{a,b}$.

It is a pleasure to thank Mark Hillery for useful discussions
and John Calsamiglia for pointing out Ref.\cite{calsa}.

\end{document}